\documentclass[12pt,a4paper]{article}
\pdfoutput=1
\usepackage{cite,color,graphics,amsmath,epsfig,rotating}
\usepackage{latexsym}
\usepackage{amssymb}

\usepackage{graphicx}
\usepackage{subfig}
\usepackage{psfrag}
\usepackage{mathrsfs}
\usepackage{url}
\usepackage[toc,page]{appendix}
\ifx\pdfoutput\undefined
\usepackage[bookmarks]{hyperref}	
\else
\usepackage{hyperref}	
\fi
\hypersetup{colorlinks=true,linktocpage,bookmarksopen,bookmarksnumbered,citecolor=azur,
linkcolor=green1,pdfstartview=FitH,urlcolor=darkred,breaklinks=true,
pdftitle={Limits on dark matter proton scattering from neutrino telescopes using micrOMEGAs},
pdfauthor={G. B\'elanger, J. Da Silva, T. Perrillat-Bottonet, A. Pukhov},
pdfsubject={paper},
pdfkeywords ={Dark matter, IceCube, neutrino}}

\usepackage{multicol}
\usepackage{color}
\definecolor{azur}{rgb}{0.118,0.498,0.796}
\definecolor{darkred}{cmyk}{0,1,1,0.4}
\definecolor{green1}{rgb}{0.21,0.6,0.32}
\def\mhref#1{\href{mailto:#1}{#1}}		

\usepackage{marginnote}

\usepackage{eurosym}
\begin{document}

\setlength{\unitlength}{1mm}
\renewcommand{\arraystretch}{1.4}
\newcommand{\comment}[1]{}


\def\micro{{\tt micrOMEGAs\;}}
\def\wimpsim{{\tt WimpSim\;}}
\def\pppc{{\tt }PPPC4DM$\nu$\;}
\def\pppcold{{\tt }DM$\nu$\;}
\def\chep{{\tt CalcHEP\;}}
\def\lhep{{\tt LanHEP\;}}
\def\darksusy{{\tt DarkSUSY\;}}

\def\ra{\rightarrow}
\def\Omg{\Omega h^2}
\def\sip{\sigma^{SI}_{\chi p}}
\newcommand{\scs}{\scriptscriptstyle}
\def\simleq{\stackrel{<}{\scs \sim}}
\def\simgeq{\stackrel{>}{\scs \sim}}
\newcommand{\com}{\textcolor{green}}
\newcommand{\combis}{\textcolor{magenta}}

\newcommand{\ablabels}[3]{
  \begin{picture}(100,0)\setlength{\unitlength}{1mm}
    \put(#1,#3){\bf (a)}
    \put(#2,#3){\bf (b)}
  \end{picture}\\[-8mm]
}

\begin{titlepage}
\begin{center}
\vspace*{-1cm}
\begin{flushright}
LAPTH-040/15\\
MAN/HEP/2015/12\\
MCnet-15-19
\end{flushright}

\vspace*{1.6cm}
{\Large\bf  Limits on dark matter proton scattering from neutrino telescopes using micrOMEGAs} 

\vspace*{1cm}\renewcommand{\thefootnote}{\fnsymbol{footnote}}

{\large 
G.~B\'elanger$^{1}$\footnote[1]{Email: \mhref{belanger@lapth.cnrs.fr}},
J.~Da Silva$^{2}$\footnote[2]{Email: \mhref{dasilva@lapth.cnrs.fr}},
T.~Perrillat-Bottonet$^{1}$\footnote[3]{Email: \mhref{thomas.perrillat-bottonet@polytechnique.edu}},
A.~Pukhov$^{3}$\footnote[4]{Email: \mhref{pukhov@lapth.cnrs.fr}},
} 

\renewcommand{\thefootnote}{\arabic{footnote}}

\vspace*{1cm} 
{\normalsize \it 
$^1\,$\href{http://lapth.cnrs.fr}{LAPTH}, Universit\'e Savoie Mont Blanc, CNRS,\\ B.P.110, F-74941 Annecy-le-Vieux Cedex, France\\[2mm]
$^2\,$\href{http://www.hep.man.ac.uk}{Consortium for Fundamental Physics, School of Physics and Astronomy}, \\ University of Manchester, Manchester, M13 9PL, United Kingdom\\[2mm]
$^3\,$\href{http://theory.sinp.msu.ru}{Skobeltsyn Institute of Nuclear Physics}, Moscow State University,\\ Moscow 119992, Russia\\[2mm]
}

\vspace{1cm}

\begin{abstract}
Limits on dark matter spin dependent  elastic scattering cross section on  protons derived from IceCube data are obtained for different dark matter annihilation channels
using~\micro. The uncertainty on the derived limits, estimated by using different neutrino spectra, can reach a factor two. For all dark matter annihilation channels except for quarks, the limits on the spin dependent cross section are more stringent than those obtained in direct detection experiments.  The new functions that allow to derive those limits are described. 

\end{abstract}

\end{center}
\end{titlepage}

\tableofcontents

\section{Introduction}

The strong astrophysical and cosmological evidence for dark matter (DM) motivates numerous direct and indirect searches for DM  both in astroparticle experiments and at colliders. 
Exploiting eventual signals from different sources could shed light on the true nature of DM. 
Indirect searches for  DM  annihilation in the galaxy are actively pursued using either positrons or antiprotons~\cite{Accardo:2014lma}, photons~\cite{Ackermann:2015zua} and neutrinos~\cite{Aartsen:2013mla,Aartsen:2014hva}. 
The best limit on galactic neutrinos have been achieved by IceCube~\cite{Aartsen:2014hva}, however the limit on the DM annihilation cross section are  still orders of magnitudes above the canonical cross section required to achieve the relic density assuming a standard cosmological scenario. 
Neutrino telescopes such as Super-Kamiokande~\cite{Tanaka:2011uf}, Baksan~\cite{Boliev:2013ai}, Amanda and IceCube~\cite{IceCube:2011aj} can also observe neutrinos originating from   
annihilation of  dark matter captured in the Sun.  In that case, the neutrino flux is determined by the cross section for DM scattering on nuclei which drives the capture rate and is thus related to direct detection searches. Neutrino telescopes are sensitive to both DM nuclei spin dependent (SD) and spin independent (SI) interactions.
However  SI interactions are currently strongly constrained by direct detection experiments such as LUX~\cite{Akerib:2013tjd} and XENON~\cite{Aprile:2012nq} whereas there is much more freedom for the SD case~\cite{Amole:2015lsj}. Moreover the  coherent enhancement of SI interactions only occurs for heavier nuclei which are subdominant in the Sun.
The prospects  to constrain  the parameter space of  DM models such as the minimal supersymmetric standard model (MSSM) using the neutrino flux from DM capture in the Sun with IceCube was examined in~\cite{Silverwood:2012tp}.
Furthermore  limits on the spin dependent cross section based on IceCube79 results have been compared to those obtained at the LHC from monojet searches  within the framework of effective field theory for a choice of DM -- quark effective operators~\cite{Blumenthal:2014cwa}, neutrino telescopes were shown to be more sensitive than colliders for heavy DM masses.

In this article we present new features of \micro~\cite{Belanger:2006is,Belanger:2013oya,Belanger:2014vza} that allow to derive the bounds on spin dependent interactions of DM with protons using the data from IceCube22, and to obtain a likelihood that allows to combine these results with those from other searches when scanning over the parameter space of a generic  DM model.  
We derive the bounds on spin dependent interactions with protons for any DM annihilation channel in standard model (SM) particles  and compare  these results with those obtained by the IceCube collaboration in two specific channels. We use the publicly available data of IceCube22, which gives a limit better than expected from statistics only because of a deficit in the number of observed events at small angles. For DM masses above ~250GeV , the limits thus obtained are expected to be comparable 
to those that can be reached using the newer data from IceCube79~\cite{Aartsen:2012kia}. However for lighter DM, the lower energy threshold of IceCube and DeepCore leads to a  higher sensitivity than IceCube22.  In the case where DM annihilates into quarks, we compare our results with those derived from monojet searches at the LHC and interpreted within an effective field theory approach assuming the same spin dependent  effective operator used for  DM capture. We furthermore examine the impact of IceCube22 data on the parameter space of dark matter models, for this we use a simple Z-portal model as well as U(1) extensions of the MSSM.

This article is organized as follows. Section~\ref{neutrino} summarizes the equations used for computing the neutrino flux, section~\ref{sec:ns} describes the statistical analysis used to extract a limit.  Section~\ref{sec:SDchannels} presents the results for the upper limit on the spin dependent cross section for each individual channel.  Section~\ref{sec:DMmodels} shows the impact of IceCube results on the parameter space of two sample dark matter models.  Our conclusions are summarized in section~\ref{sec:Conclusion}. The \micro functions are described in the appendix as well as the method used to compare our results with \darksusy~\cite{Gondolo:2004sc}.

\section{Neutrino Flux}
\label{neutrino}

After being captured, DM particles concentrate in the center of the Sun and 
then  annihilate into Standard Model particles. These SM particles further decay producing neutrinos that can be 
observed at the Earth.  The equation describing the evolution of the number of DM particles $N_\chi$  (assuming the DM is self-conjugate) reads
 \begin{eqnarray}
   \dot{N}_\chi& =& C_\chi -A_{\chi\chi} N_\chi^2 -E N_\chi\,, 
\label{eq:ndot}
\end{eqnarray}
where  $A_{\chi\chi}$ is the rate of depletion of DM,  
\begin{equation}
A_{\chi\chi}=  \frac{{\langle \sigma v \rangle}_{\chi\chi}}{V_{eff}}\,,
\label{eq:ann}
\end{equation}
  $\langle\sigma v\rangle_{\chi\chi}$ is the velocity averaged annihilation cross section of DM into SM particles and  $V_{eff}$ is the effective volume of DM in  the Sun. $E$ is the evaporation rate and $C_\chi$ is the capture rate of DM particles in the core of the Sun. It  depends on the DM -- nucleus scattering  cross section, as well as on the DM velocity distribution and local density~\cite{Gould:1987ir}.   
 The computation of the capture rate in \micro~is described in~\cite{Belanger:2013oya} and involves a contribution from spin independent and spin dependent interactions  of DM on nuclei.
The spin independent interactions  add coherently,  so the contribution of heavy nuclei is enhanced with a factor  $\propto A^2$. However spin independent interactions are strongly constrained from direct detection experiments and  therefore can safely be neglected as will be demonstrated in section ~\ref{sec:SDchannels}.  For spin dependent interactions there is no coherence effect and hydrogen gives the largest contribution, neutrino telescopes are therefore mostly sensitive to spin dependent interactions on protons.

Neglecting the evaporation rate, the solution of Eq.~\eqref{eq:ndot} is simply 
\begin{equation}
\label{eq:capture}
\Gamma_{\chi\chi}=\frac{1}{2} A_{\chi\chi} N_\chi^2= \frac{C_\chi}{2} \tanh^2(\sqrt{C_\chi A_{\chi\chi}}t)
\end{equation}
where $t=4.57\times 10^9$ yrs is the age of the Sun. When the capture rate and the annihilation cross section are sufficiently large ($\sqrt{C_\chi A_{\chi\chi}} t \gg 1$) equilibrium is reached and  the annihilation rate only depends on the capture rate,  $\Gamma_{\chi\chi}=C_\chi/2$  and not on DM annihilation properties.

The total neutrino (antineutrino) spectrum at the Earth is proportional to the capture rate but also features a dependence  on the DM annihilation channel
\begin{eqnarray}
  \label{eq:nu_spectra}
  \frac{d\phi_\nu}{dE_\nu} &=& \frac{1}{4\pi d_\odot^2} 
             \Gamma_{\chi\chi} \sum_f Br_{f\bar{f}}\, \frac{dN_f}{dE_\nu}\,
\end{eqnarray}
where $d_{\odot}=1.5\times 10^8$~km is the distance to the Sun, 
$Br_{f\bar{f}}$ the branching fraction into each particle/antiparticle final state $f\bar{f}$ and
$N_f$  are the neutrino spectra resulting from  those annihilations.
The neutrino spectra originating from different annihilation channels into SM particles and 
taking into account oscillations  and Sun medium effects were  computed in
\wimpsim~\cite{Blennow:2007tw}, \pppcold~\cite{Cirelli:2005gh} and \pppc~\cite{Baratella:2013fya}\footnote{Note that neutrino signals from the Sun and corresponding limits on spin dependent interaction cross sections were also investigated in \cite{Ibarra:2013eba} by focusing on the effect of internal bremsstrahlung and in \cite{Ibarra:2014vya} by paying attention on consequences of final state radiation of electroweak gauge boson.}. We use the set of tables provided by these groups. 
 
  The distribution of the number of events in the detector, $N_s$,  will depend on the flux of muon neutrinos.  After an exposure time $t_e$ the number of events  is given by 
  \begin{equation}
  \label{eq:dNsdE}
\frac{dN_s}{dE}=t_e \left(\frac{d\phi_{\nu_{\mu}}}{dE}A_{\nu}(E) + \frac{d\phi_{\bar{\nu}_{\mu}}}{dE}A_{\bar{\nu}}(E)\right)
  \end{equation}
where $A_\nu(E)$ and $A_{\bar\nu}(E)$ are the effective area of the detector for neutrinos and anti-neutrinos respectively.

\section{Limits on the number of signal events}
\label{sec:ns}

To obtain the limits on the neutrino flux that could originate from DM capture in the Sun, one first needs to  distinguish the signal  and background events. 
The neutrino fluxes depend on the angle $\varphi$ between the incoming neutrino  and the direction of the Sun. 
Background events are expected to be distributed over any angle $\varphi$ whereas all signal events should occur at  $\varphi=0$. 
However the IceCube experiment has a finite angular resolution, the angular distribution of signal events is therefore
\begin{equation}
\label{eq:N_cosphi}
   \frac{dN}{d\cos{\varphi}}= C  e^{\frac{ \cos{\varphi}-1}{\sigma^2}},
\end{equation}
where $C$ is a normalization constant and $\sigma$  is the mean angle (in radians)~\cite{Scott:2012mq}. 
The mean and median angle, $\varphi_{med}$ (in degrees) are simply related through 
\begin{equation}
  \sigma=\frac{\varphi_{med}}{\sqrt{2\log(2)}}\frac{\pi}{180}.
\end{equation}
 The energy dependence of the median angle  is provided on the IceCube official website\footnote{\href{https://icecube.wisc.edu/science/data/ic22-solar-wimp}{https://icecube.wisc.edu/science/data/ic22-solar-wimp}}.
For example, for a  DM of mass 1000~GeV which annihilates into $W^+W^-$,  $\varphi_{med}=2.9^{\circ}$ \cite{Abbasi:2009uz}. 

After estimating the background  using neutrino events registered at large angles, an upper limit on the number of signal events can be obtained from the number of events registered at small angle, this method allows to get rid of systematic uncertainties.  
To obtain an upper limit on the number of signal events, IceCube22 uses the Feldman-Cousins method \cite{Feldman:1997qc}. 
In IceCube22, 13 events were registered in the  angular bin $\varphi<3^\circ$  while 18 background events are expected.   
From the number of events in that angular bin we have checked that the   upper limits on the total number of signal events, $\mu_{90}$, agree with the ones presented in ~\cite{Abbasi:2009uz} and displayed in  the second column of Table~\ref{tab:signalToCS} for the case of DM annihilating into $W^+W^-$.

\begin{table}[htbp]
\begin{center}
\begin{tabular}{c|c|cccc|c}
\hline \hline
 $m_\chi$&$\mu_{90}$\cite{Abbasi:2009uz} &{\footnotesize \micro}  &{\footnotesize \micro }&{\footnotesize \darksusy} &{\footnotesize IceCube22} &$\mu_{90}^{micrO}$ \\
 & &{\footnotesize \pppc} &\footnotesize{\wimpsim}  & & &\\
\hline \hline
 250 & 7.5     &$1.37\;10^{-40}$ & $9.51\;10^{-41}$     &$8.9\;10^{-41} $  &$2.8\;10^{-40}$ & 9.07\\
 500 & 6.8     &$1.48\;10^{-40}$&$1.04\;10^{-40}$       &$9.7\;10^{-41} $&$3.0\;10^{-40}$   & 8.98\\
1000 & 6.8     &$4.38\;10^{-40}$ &$2.67\;10^{-40}$      &$2.5\;10^{-40} $&$8.7\;10^{-40}$   & 8.95\\
3000 & 6.4     &$3.00\;10^{-39}$ &$2.06\;10^{-39}$      &$2.1\;10^{-39} $&$9.9\;10^{-39}$   & 8.65\\
5000 & 6.8     &$7.84\;10^{-39}$  &$6.00\;10^{-39}$     &$5.8\;10^{-39} $&$3.6\;10^{-38}$   & 8.67\\
\hline \hline
\end{tabular}
 \caption{  
90\%CL upper limit on the  number of signal events, $\mu_{90}$, for DM of mass $m_\chi$ in [GeV] annihilating into  $W^+W^-$ and the corresponding limits in [cm$^2$] on spin dependent DM -- proton cross section obtained by \micro with 
\pppc or \wimpsim, by \darksusy, and by IceCube~\cite{Abbasi:2009uz}. The last column contains the 90\%CL upper limit on the number of signal events $\mu_{90}$ evaluated using the likelihood function of  \micro. 
}
\label{tab:signalToCS}
\end{center}
\end{table}

An alternative approach to derive the upper limit on the number of events based on Bayesian statistics has been implemented in \micro. The likelihood function uses 
 both information about the angular distribution and the number of digital optical modules (DOM) activated for each event. Note that the number of active DOMs is correlated with neutrino energy.
The method makes use of the  experimental data provided by the IceCube22 collaboration  and  in the DarkSUSY-5.1.2 distribution
\cite{Scott:2012mq}\footnote{shared/DarkSusy/IC\_data}, this includes,

\begin{itemize}
\item $A_{\nu}(E)$ - the  effective area for neutrino detection;
\item $A_{\bar{\nu}}(E)$ - the effective area for anti-neutrino detection;
 \item $\sigma(E)$ - the mean value of the neutrino angle resolution;
 \item $ch_k(N_{chan})$ - the probability to activate $N_{chan}$ DOMs~\cite{Scott:2012mq}. This is given for 20 energy intervals in the range 
$39$~GeV~$<E_\nu<4\;10^5$~GeV;
\item  $\frac{dN_{bg}}{d\cos{\varphi}}$ - the distribution of background events as a function of the angle between  the incoming (anti)neutrino and the Sun;   
\item  $ch_{bg}(N_{chan})$ - the probability to activate $N_{chan}$ DOMs for background event;
\item  $t_e $ the exposure time  $t_e=104/365$ [Year].   
\end{itemize}

For background events, the  angular and energy distribution are not expected to be correlated. Then the probability of observing a  background event with an angle $\varphi$ that
activates  $N_{chan}$ DOMs in the detector  is 
\begin{equation}
    \mathbb{P}_{bg}( \cos\varphi,N_{chan})= \frac{dN_{bg}}{d\cos{\varphi}} ch_{bg}(N_{chan}) \,.
\end{equation}

The probability of a signal event  for given neutrino $\phi_{\nu_{\mu}}$ and  anti-neutrino $\phi_{\bar{\nu}_{\mu}}$ fluxes is defined as
\begin{equation}  
\label{eq:LLsignal}
  \mathbb{P}_s(\cos\varphi,N_{chan})= \sum_{k=1}^{20} \frac{n_k} {\sigma_k^{2}} e^{\frac{(\cos\varphi-1)}{\sigma_k^2}} ch_k(N_{chan})
\end{equation}
where 
\begin{eqnarray} 
n_k&=&       \int_{E_{k-1}}^{E_k} \frac{dN_s}{dE}dE \\ 
\sigma_k^2&=&\frac{1}{n_k}\int_{E_{k-1}}^{E_k} \frac{dN_s}{dE}\sigma(E)^2dE \,.
\end{eqnarray}

The IceCube collaboration also provides the  full list of events registered  during the  exposure time $t_e$, altogether 6946 events.
Most of these events occur at large angle with respect to the direction of the Sun. 
To construct a  likelihood function we  avoid the regions with a large atmospheric  background, and we will consider only events at angles less than  $\varphi_{cut}$. The number of expected events for both background ($N_{bg}$) and  signal ($N_s$) is then
\begin{eqnarray}
N_{bg}^c&=&  \sum_{N_{chan}} \int\limits_0^{\cos{\varphi_{cut}}} \mathbb{P}_{bg}(\cos\varphi,N_{chan}) d\cos\varphi\\
N_s ^c  &=&  \sum_{N_{chan}}  \int\limits_0^{\cos{\varphi_{cut}}} \mathbb{P}_{s}(\cos\varphi,N_{chan}) d\cos\varphi \,.
\end{eqnarray} 
The likelihood function reads 
\begin{equation}
\label{eq:likelihood}
{\cal L}(\alpha,\varphi_{cut})=\frac{ (N_{bg}^c +\alpha N_s^c)^{N_{tot}}}{N_{tot}!} e^{-N_{bg}^c-\alpha N_s^c} \prod_{i=1}^{N_{tot}} \mathbb{P}(\cos\varphi_i,N_{{chan},i},\alpha)
\end{equation}
where  $N_{tot}$  is the  total number of detected events for $\varphi<\varphi_{cut}$. The coefficient $\alpha$ varies in the range
$[0,\infty]$  and
 \begin{equation}
\mathbb{P}(\cos\varphi,N_{chan},\alpha)=  \frac{1}{N_{bg}^c+\alpha N_s^c} \left(   \mathbb{P}_{bg}(\cos\varphi,N_{chan}) + \alpha  \mathbb{P}_{s}(\cos\varphi,N_{chan}) \right) \,.
\end{equation} 
To obtain  the limit on the number of signal events we   use  the Bayesian
approach with a flat prior distribution for the number of signal events. Then the
confidence level or rather {\it credibility} level, is given by~\cite{Agashe:2014kda}
\begin{equation}
\label{eq:pvalue}
  CL=  {\int\limits_0^1{\cal L}(\alpha) d\alpha}/
                         {\int\limits_0^\infty {\cal L}( \alpha) d\alpha} \,,
\end{equation}
Here CL  represents the probability that the number of signal events is less than $N_s^c$. 

\begin{figure}[h!]
\begin{center}
\vspace{-0.3cm}
\hspace{-12mm}\includegraphics[scale=0.8]{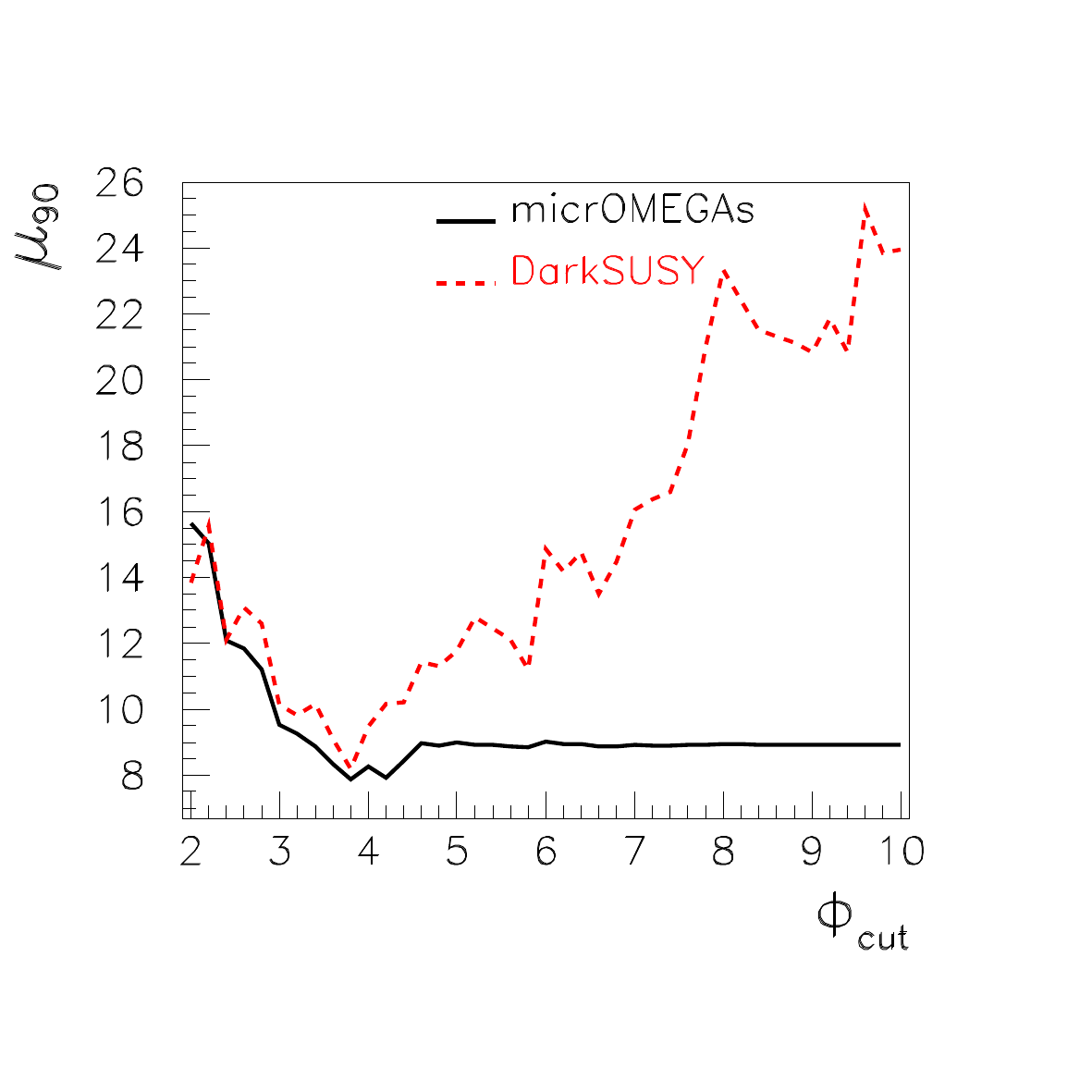}
\vspace{-1.2cm}
\caption[]{ 90\%CL limit on the total number of signal events as a function of $\varphi_{cut}$  for a DM mass of 1~TeV in the $W^+W^-$ channel obtained with   \micro (black) and \darksusy (dash/red). 
}
\label{fig:cut}
\end{center}
\end{figure}

Fig.~\ref{fig:cut} shows the  dependence of the 90\%CL upper limit on  the number of signal events
 on $\varphi_{cut}$. Clearly the limit does not depend on  $\varphi_{cut}$  for  $\varphi_{cut} > 5^{\circ}$,  
in our calculations we fix $\varphi_{cut} =10^{\circ}$. 
For comparison the dependence  on $\varphi_{cut}$  of the upper limit on the number of signal events obtained with the default option in \darksusy  is also displayed. Note that the two methods give similar results for
$\varphi_{cut}=3.7^{\circ}$, this angle also corresponds to the maximal value for $N_s^c/\sqrt{N^c_{bckg}}$.
The  90\%CL limits on the number of signal events  calculated by \micro are presented in the last column of Table~\ref{tab:signalToCS}.  Despite the different statistical methods used, these numbers are in rough agreement with those extracted using only
one  angular bin as done by IceCube22~\cite{Abbasi:2009uz}.

Note that to take into account the 20\% uncertainty on the effective area, {\tt micrOMEGAs}
divides $A_{\nu}(E)$, $A_{\bar{\nu}}(E)$ by $1.2$ when  calculating the CL. 
Therefore we  effectively rescale by 20\% the limit  extracted on the SD cross section and we thus obtain a more conservative limit. Note that this rescaling has no impact on the upper limit on the number of signal events.

\section{Extracting limits on DM -- proton spin dependent cross sections}
\label{sec:SDchannels}

For a given number of signal events, one can extract an upper limit on the DM -- nuclei cross section, this limit   will  depend on the fluxes of muon neutrinos/antineutrinos  that reach the
detector, Eq.~\eqref{eq:nu_spectra}. This spectrum will depend primarily on the DM annihilation channel. When neutrinos propagate through the Sun,  the spectra will be distorted by
neutrino attenuation in the Sun and neutrino oscillation, furthermore other effects like electroweak bremstrahlung can also play a role.
The latter is particularly important for the electron and muon channels, otherwise these channels would not lead to neutrinos because the electron is stable and the muon is quickly absorbed in the Sun.
Two sets of neutrino spectra  are publicly available, \wimpsim\cite{Blennow:2007tw} and  \pppc
\cite{Baratella:2013fya}. Both  spectra  are implemented in~\micro and are used  to compute the neutrino flux.  
We have checked the validity of our implementation by comparing with the spectra presented in \cite{Blennow:2007tw, Baratella:2013fya} and directly with the \wimpsim tables  implemented in  DarkSUSY-5.1.2~\cite{Gondolo:2004sc}.  
Furthermore we have also compared our current implementation with the  previous version of neutrino spectra \pppcold\cite{Cirelli:2005gh}
implemented in micrOMEGAs\cite{Belanger:2013oya}.

\begin{figure}[h!]
\begin{center}
\vspace{-1cm}
\begin{tabular}{ccc}
\hspace{-12mm}\includegraphics[scale=0.4]{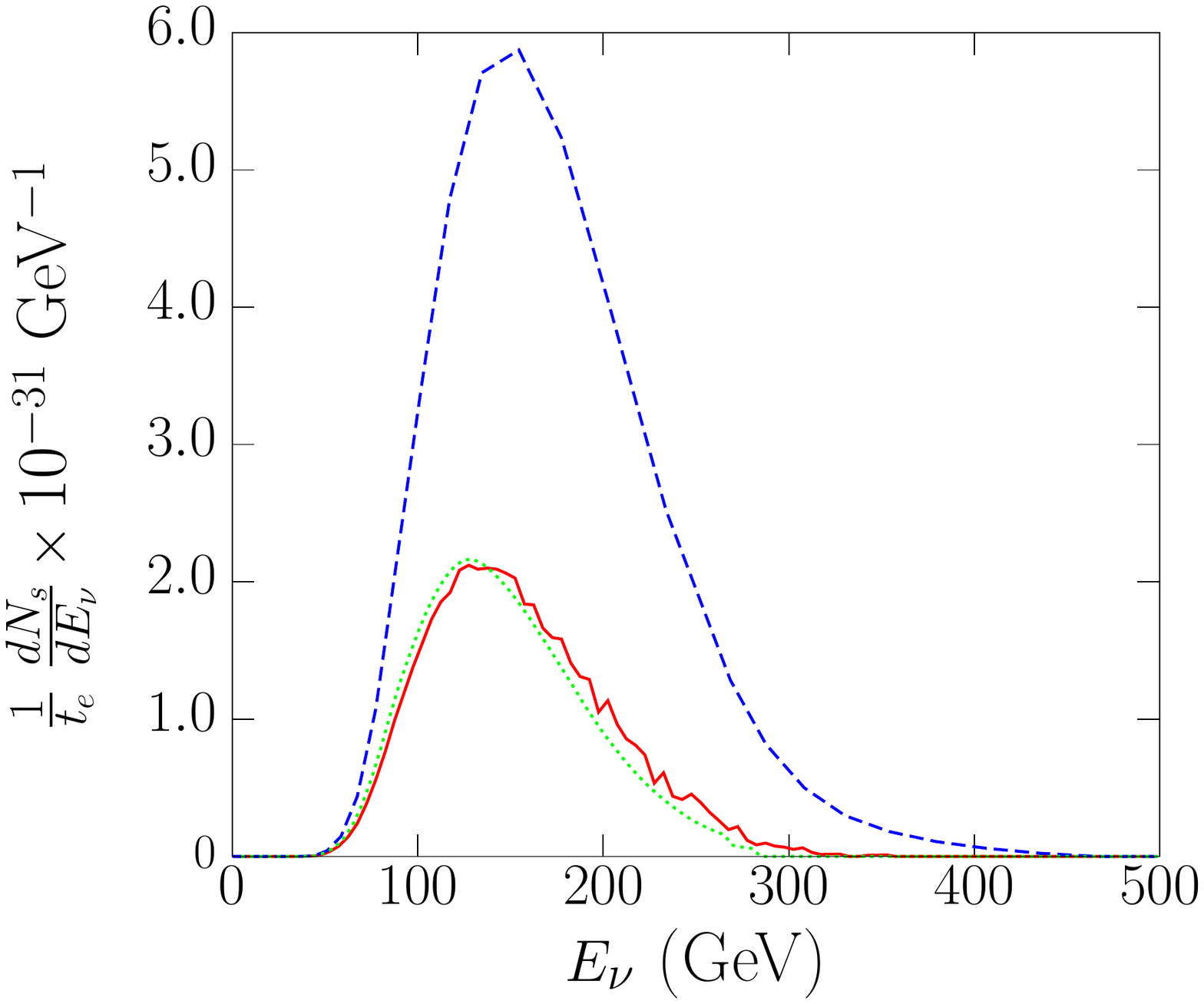}&
\hspace{-12mm}\includegraphics[scale=0.4]{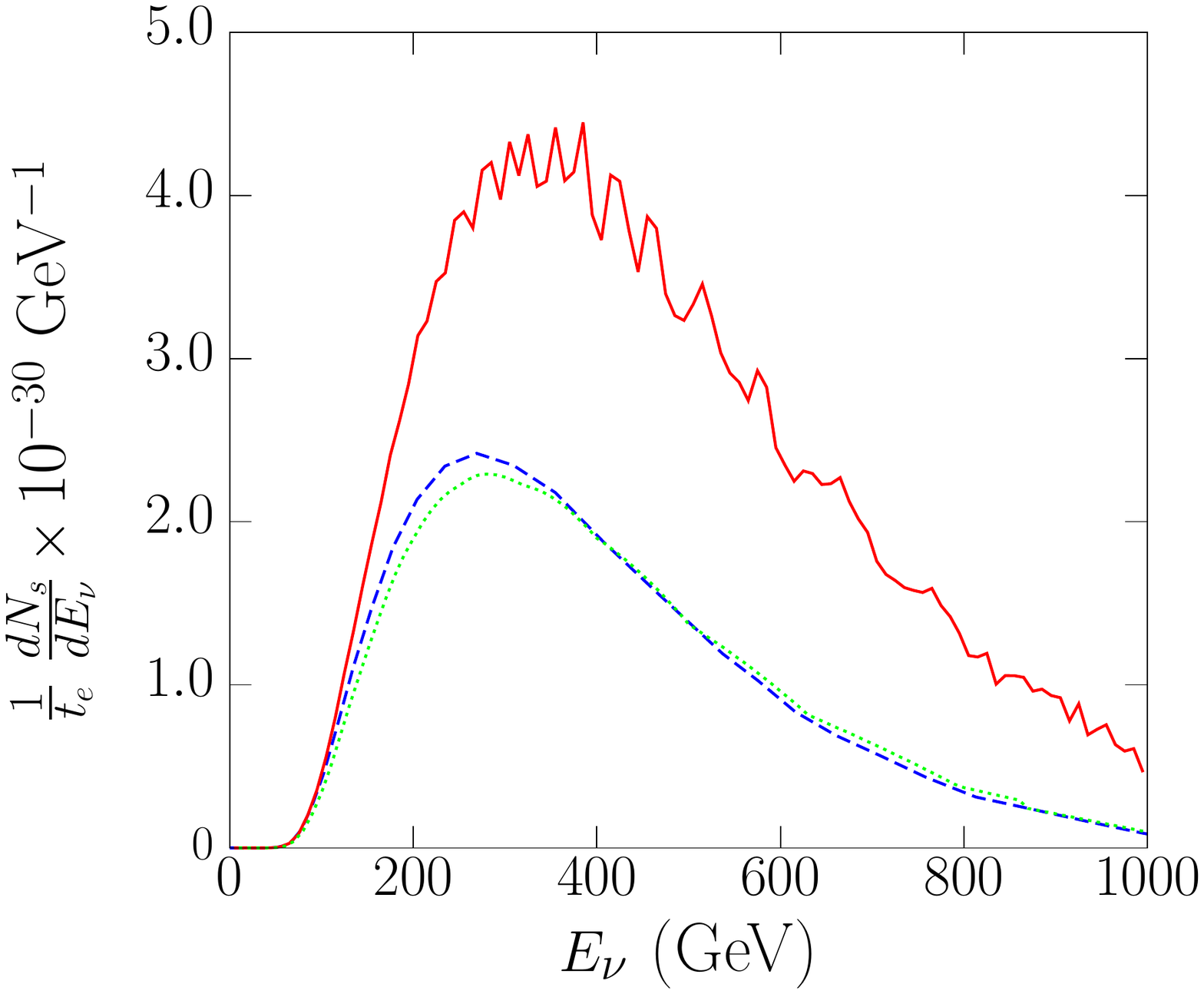}
\end{tabular}
\vspace{-2.8cm}
\caption[]{ Comparison of neutrino spectra from  \pppc (dashed/blue) and  \wimpsim (full/red) for DM annihilation into $b\bar{b}$ (left) and  $W^+W^-$ (right).
The DM mass is set to 500~GeV for $b\bar{b}$ and 1~TeV for $W^+W^-$. The spectra from the \pppcold tables~\cite{Cirelli:2005gh} are also displayed for comparison (dotted, light green).}
\label{fig:comparison}
\end{center}
\end{figure}
 
For some channels the  \wimpsim and \pppc spectra can be quite different, see Fig.~\ref{fig:comparison} which compares the total neutrino/anti-neutrino spectra for two different channels. This difference will impact
 the limits derived  on the DM -- nucleon cross sections. 
The reason for these  differences is not completely  clear despite the different approaches used. \wimpsim uses Pythia~\cite{Sjostrand:2007gs} to get the initial neutrino spectra and  a Monte Carlo approach for neutrino propagation
while \pppc uses GEANT~\cite{Agostinelli:2002hh} for the initial spectra and solves the equation for the density matrix.
The discrepancy in the $b\bar{b}$ channel could be due to differences  in Pythia and GEANT since the previous version of the neutrino spectrum \pppcold\cite{Cirelli:2005gh}, which relied on Pythia is in perfect agreement with \wimpsim.  
We have also noted that the different treatment of the propagation in the Sun leads to different neutrino spectra even in the case where DM annihilates directly into neutrino pairs, only the \wimpsim spectrum featuring a peak at the maximum energy. 
In such case it is not surprising that for all channels with an initial hard neutrino spectrum, for example $W^+W^-$, \wimpsim leads to a larger and harder flux, see the right panel of Fig.~\ref{fig:comparison}.
Another difference between the two codes is that  bremstrahlung is included in \pppc, however bremstrahlung is expected to be important mainly for light leptons.

The neutrino spectra produced by polarized vector bosons and  leptons are also provided in 
\pppc. 
The polarisation of vector bosons affects the  spectrum of high energy neutrinos at the source, after propagation the effect is of the order of 10\%.
Although \micro has an option to automatically compute the polarization of vector bosons in DM annihilation processes, the neutrino signal in a given model is always computed assuming unpolarized vector bosons since the effect is not large. The polarized tables can be used only in a model independent approach when one inputs a cross section and chooses explicitly the annihilation channel. 
On the other hand~\micro does not keep track of the  polarization of leptons. The polarization effect can be much more important for
annihilation channels into light leptons, with in particular an important contribution from left-handed light leptons ($e,\mu$) due to bremstrahlung. However  the neutrino flux  is expected to 
be very small for these channels as will be demonstrated in the next section. Furthermore the neutrino spectra only depend weakly on the $\tau$ polarization.

\subsection{Comparison with \darksusy}

The value of the DM -- proton cross section corresponding to the upper limit on the number of signal events follows from Eqs.~\eqref{eq:capture}, \eqref{eq:nu_spectra} and \eqref{eq:dNsdE}.
The results computed with \micro~and using two different choices of neutrino spectra \cite{Blennow:2007tw}, \cite{Baratella:2013fya} are  also presented in Table~\ref{tab:signalToCS}. 
These values are obtained starting with the upper limit on the number of events given in the first column and obtained in ~\cite{Abbasi:2009uz}.
The results obtained with the \wimpsim spectra are compared with those obtained with \darksusy\cite{Gondolo:2004sc} and are found to be in excellent agreement -- 
for a description of the method used see appendix~\ref{usageDS}.  
The limits presented by the IceCube collaboration on the spin dependent cross section  are higher by roughly a factor of three~\cite{Abbasi:2009uz} and are 
 consistent with the ones obtained using the  \darksusy option to compute the 90\%CL 
using  $\varphi_{cut}=8^\circ$. 
Note that the limits obtained with \darksusy strongly depend on the value chosen for the angle.
Since  the signal is concentrated at small angle, 
the   limits can be improved with a smaller angular cut.
To first approximation, the optimal cut corresponds  to the maximum  of  $N_{signal}(\phi_{cut})/\sqrt{N_{bg}(\phi_{cut})}$. 
For a  1~TeV DM annihilating  into $W^+W^-$, the maximum is reached at $\phi_{cut}=3.7^\circ$.  For this cut, the  
90\% exclusion limit on the number of signal events  obtained with \darksusy
is  $8.5$, in good agreement with the \micro result shown in the last column of
Table~\ref{tab:signalToCS}.  

In Table~\ref{tab:icecube_new_W} we compare  the 90\% confidence limit on the DM-proton SD cross section obtained with \darksusy and with \micro for the case of  DM annihilation into $W^+W^-$. 
For this we use the default ``Number likelihood'' flag of \darksusy,
For  each DM mass  we find the angular cut  which minimizes the ratio
$N_{signal}(\phi_{cut})/\sqrt{N_{bg}(\phi_{cut})}$ where $N_{signal}$
is the predicted number of signal events   and $N_{bg}$ is the number
of background events estimated by \darksusy. Note that $\phi_{cut}$ 
does not depend on  the DM-proton cross section nor on data.  These results are compared with  the ones obtained with \micro using only angular distribution. 
The value obtained with \micro is larger because we allow an additional 20\% uncertainty to take into account the systematic error on the effective area of IceCube22.
We also compare with the results obtained with   \micro adding the information from the energy distribution as described in the previous section.  Here limits are  improved noticeably for DM masses below 1 TeV, while at higher masses the two methods give similar results,  this is because the energy distribution of signal and background events are similar.
The same comparison is performed for the $b\bar{b}$ channel and presented in Table~\ref{tab:icecube_new_B}.

\begin{table}[htbp]
\begin{center}
\begin{tabular}{c|ccccc}
\hline \hline
\multicolumn{1}{c|}{$m_\chi$}  & \footnotesize{IceCube22} 
&\footnotesize{\darksusy} &\footnotesize{\micro}  &\footnotesize{\micro} & \footnotesize{IceCube79}  \\
    (GeV)  &\cite{Abbasi:2009uz}  
&\footnotesize{with $\phi$
cut}&\footnotesize{ $\phi$ only}&\footnotesize{$\phi + N_{chan}$ } &\cite{Aartsen:2012kia}
 \\
\hline \hline
100   &                  & $4.19\;10^{-39}(5.6^\circ)$   & $4.65\;10^{-39}$ & $3.25\;10^{-39}$   & $2.68\;10^{-40}$\\
250   & $2.8\;10^{-40}$  & $1.33\;10^{-40}(4.3^\circ)$   & $1.56\;10^{-40}$  & $1.38\;10^{-40}$ & $1.34\;10^{-40}$ \\
500   & $3.0\;10^{-40}$  & $1.67\;10^{-40}(3.9^\circ)$     & $1.71\;10^{-40}$ & $1.64\;10^{-40}$& $1.57\;10^{-40}$ \\
1000  & $8.7\;10^{-40}$  & $3.12\;10^{-40}(3.7^\circ)$    & $4.13\;10^{-40}$ & $4.21\;10^{-40}$ & $4.48\;10^{-40}$ \\
3000  & $9.9\;10^{-39}$  & $2.73\;10^{-39}(3.7^\circ)$    & $3.32\;10^{-39}$ & $3.32\;10^{-39}$ & $5.02\;10^{-39}$ \\
5000  & $3.6\;10^{-38}$ & $7.24\;10^{-39}(3.7^\circ)$     & $9.09\;10^{-39}$ & $9.09\;10^{-39}$& $1.59\;10^{-38}$  \\
\hline \hline
\end{tabular}
 \caption{Comparison between limits on {$\sigma^{SD}_{\chi p}$ (cm$^2$) for $\chi \chi \rightarrow W^+W^-$}  
obtained by IceCube,  DarSUSY for the value of $\phi_{cut}$ given in parenthesis and \micro  using only angular information ($\phi$ only) and also energy distributions ($\phi + N_{chan}$). All results are obtained using  the \wimpsim tables. }
\label{tab:icecube_new_W}
\end{center}
\end{table}

\begin{table}[htbp]
\begin{center}
\begin{tabular}{c|ccccc}
\hline \hline
\multicolumn{1}{c|}{$m_\chi$}  & \footnotesize{IceCube22} 
&\footnotesize{\darksusy} &\footnotesize{\micro}  &\footnotesize{\micro} & \footnotesize{IceCube79}  \\
    (GeV)  &\cite{Abbasi:2009uz}  
&\footnotesize{with $\phi$
cut}&\footnotesize{ $\phi$ only}&\footnotesize{$\phi + N_{chan}$ } &\cite{Aartsen:2012kia}
 \\
\hline \hline
100    &                   & $1.63\;10^{-36}(6.2^\circ)$ & $3.74\;10^{-36}$ & $2.59\;10^{-36}$  & $1.47\;10^{-38}$    \\
250    &                  & $2.94\;10^{-38}(5.0^\circ)$ & $3.23\;10^{-38}$& $2.40\;10^{-38}$  & $5.90\;10^{-39}$     \\
500    & $2.6\;10^{-38}$   & $1.13\;10^{-38}(4.5^\circ)$ & $1.13\;10^{-38}$& $9.46\;10^{-39}$  & $7.56\;10^{-39}$   \\
1000   & $2.2\;10^{-38}$   & $1.02\;10^{-38}(4.2^\circ)$ & $1.16\;10^{-38}$& $1.05\;10^{-38}$  & $1.00\;10^{-38}$     \\
3000   & $7.2\;10^{-38}$   & $2.69\;10^{-38}(4.0^\circ)$ & $3.55\;10^{-38}$& $3.35\;10^{-38}$  & $3.16\;10^{-38}$      \\
5000   & $1.5\;10^{-37}$   & $4.75\;10^{-38}(4.1^\circ)$ & $7.43\;10^{-38}$& $7.28\;10^{-38}$  & $7.29\;10^{-38}$     \\
\hline \hline
\end{tabular}
 \caption{Same as Table~\ref{tab:icecube_new_W}    
for $\chi \chi \rightarrow b\bar{b}$.}
\label{tab:icecube_new_B}
\end{center}
\end{table}
Finally, we comment on more recent results from the IceCube79 collaboration~\cite{Aartsen:2012kia}.  The lower energy threshold achievable with the use of  DeepCore  leads to a significant  increase in  sensitivity for DM masses below $\approx 250$~GeV. At high masses 
 the limits that are obtained by the collaboration in the $W^+W^-$ and $b\bar{b}$ channel are roughly a factor 2 better than those of IceCube22~\cite{Abbasi:2009uz}, see
Tables~\ref{tab:icecube_new_W} and \ref{tab:icecube_new_B}.
 However we have already shown that the likelihood function implemented in \micro leads also to better limits than those found in~\cite{Abbasi:2009uz}. In part it is due to a statistical anomaly associated with the small  number of detected events at small angles (13 events observed and 18 events expected).
 
Note that using the \pppc spectra leads to a more conservative limit in the WW channel, this is expected considering the difference in neutrino spectra mentioned above. This is not the case in all channels as will be discussed in the next section.

\subsection{Model independent limits on DM -- proton cross section}

\begin{figure}[h!]
\begin{center}
\begin{tabular}{ccc}
\hspace{-12mm}\includegraphics[scale=0.44]{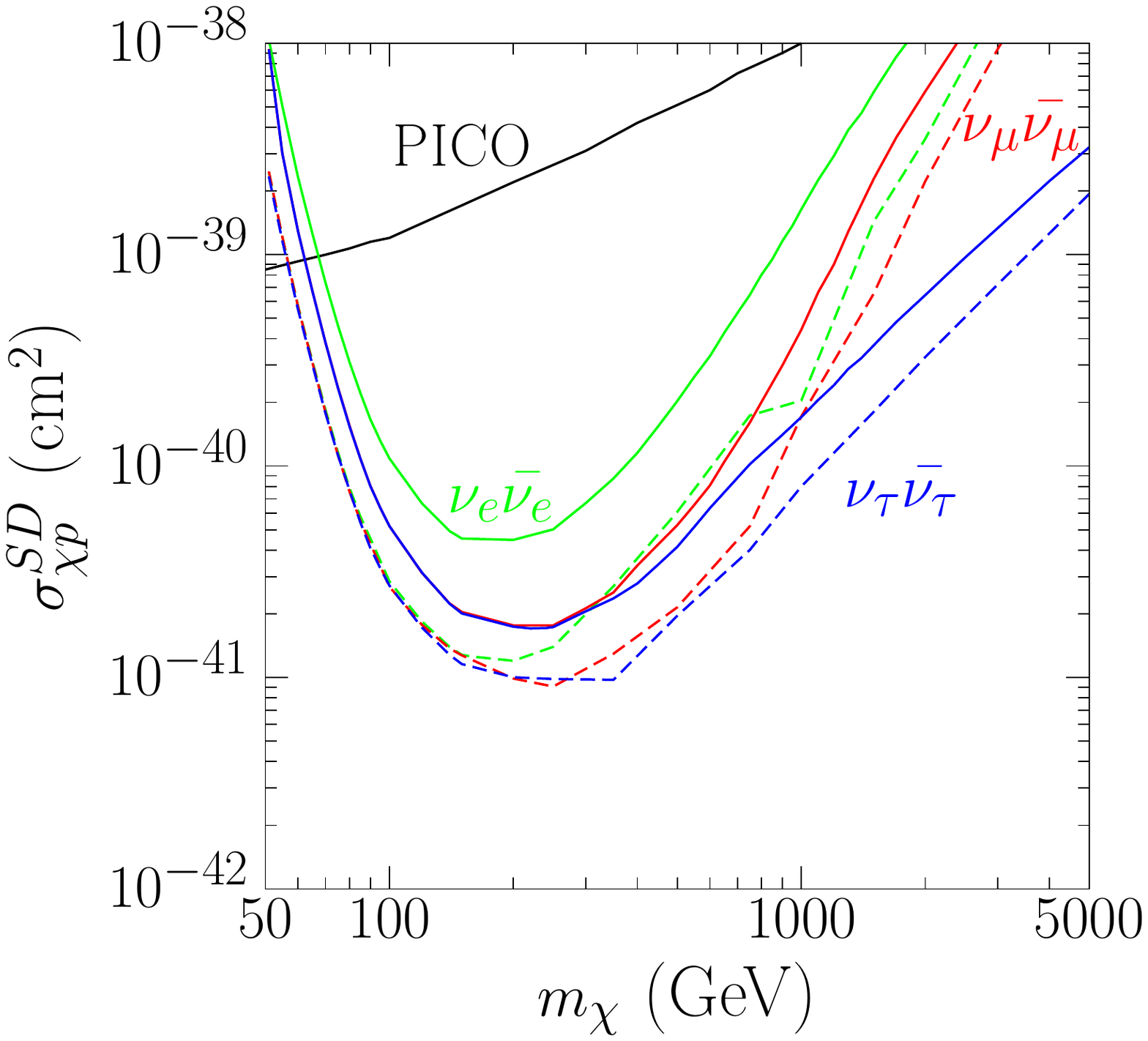}
\hspace{-12mm}\includegraphics[scale=0.44]{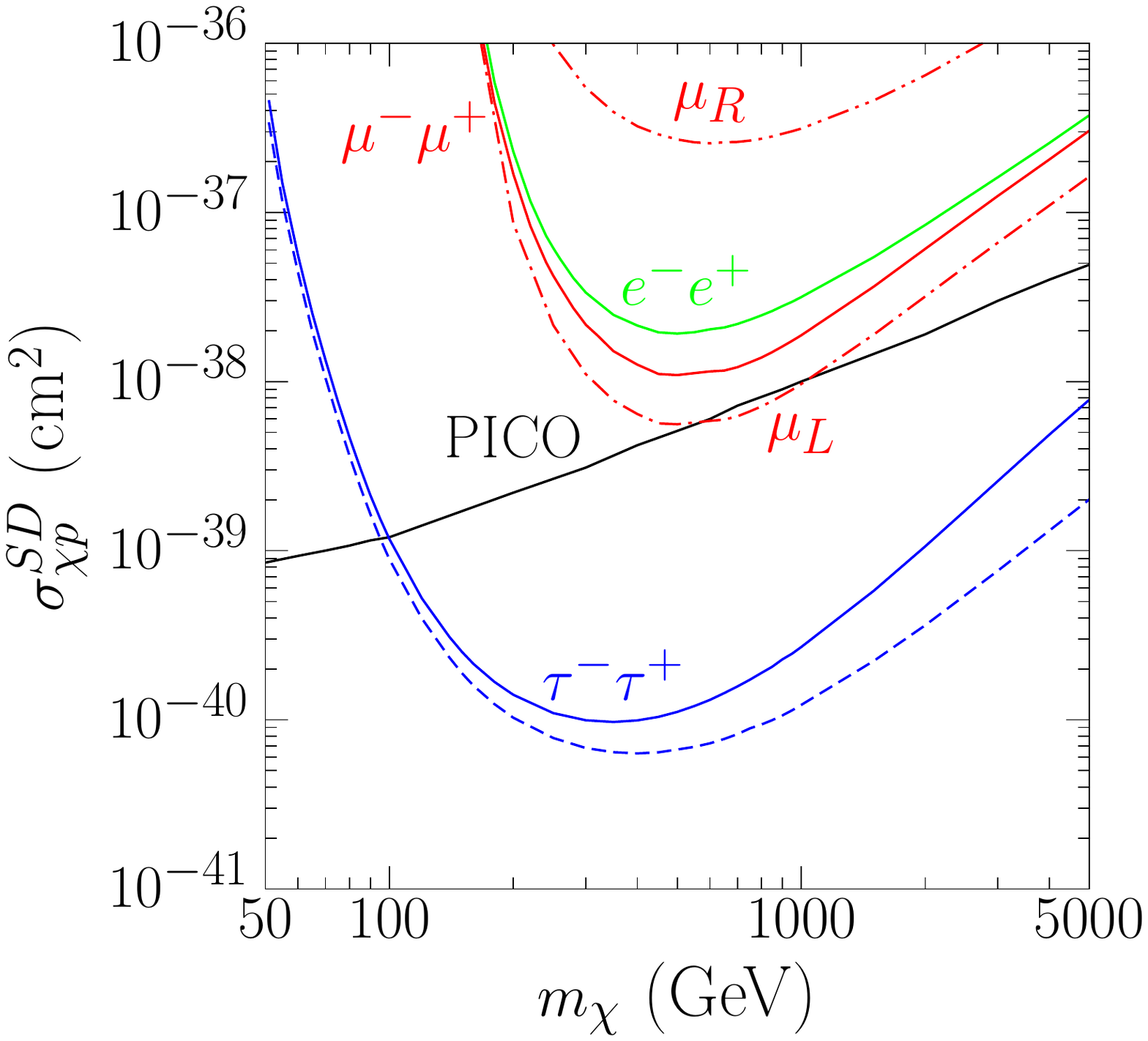}
\end{tabular}
\vspace{-2.8cm}
\caption[]{Upper limit on $\sigma^{SD}_{\chi p}$ for DM annihilation in $\nu_i\bar\nu_i$, $i=e,\mu,\tau$ (left panel) or  $\tau^+\tau^-$ (right panel) using the \pppc(full) or \wimpsim(dashed) spectra.
The best limit on  $\sigma^{SD}_{\chi p}$ from the direct detection experiment PICO~\cite{Amole:2015lsj} is shown for comparison. The right panel also shows the limits from  the light lepton  channels  using \pppc(full). The limits for polarized muons (dashed) are also displayed.}
\label{fig:leptons}
\end{center}
\end{figure}

\begin{figure}[h!]
\begin{center}
\begin{tabular}{ccc}
\hspace{-12mm}\includegraphics[scale=0.44]{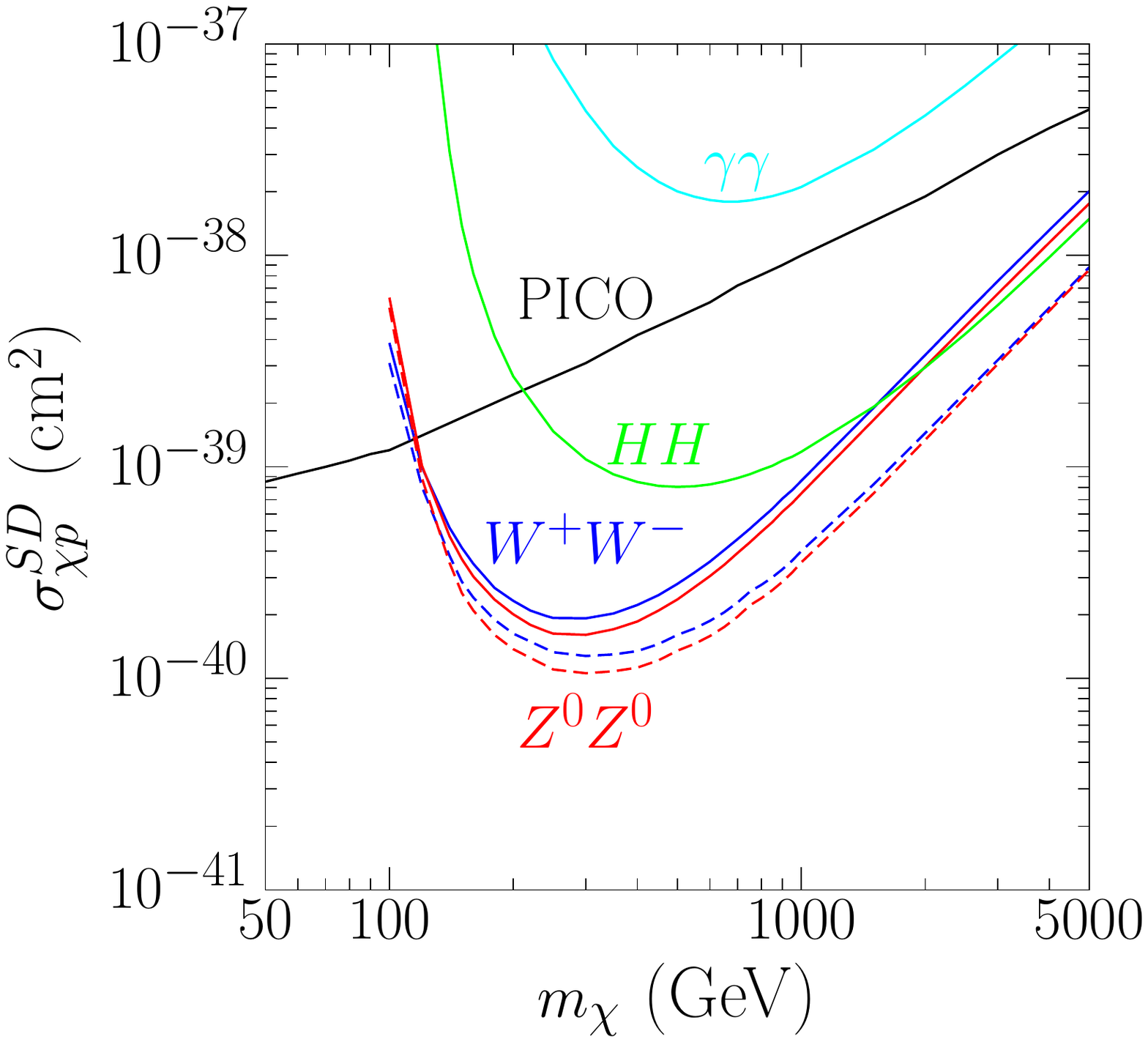}
\hspace{-12mm}\includegraphics[scale=0.44]{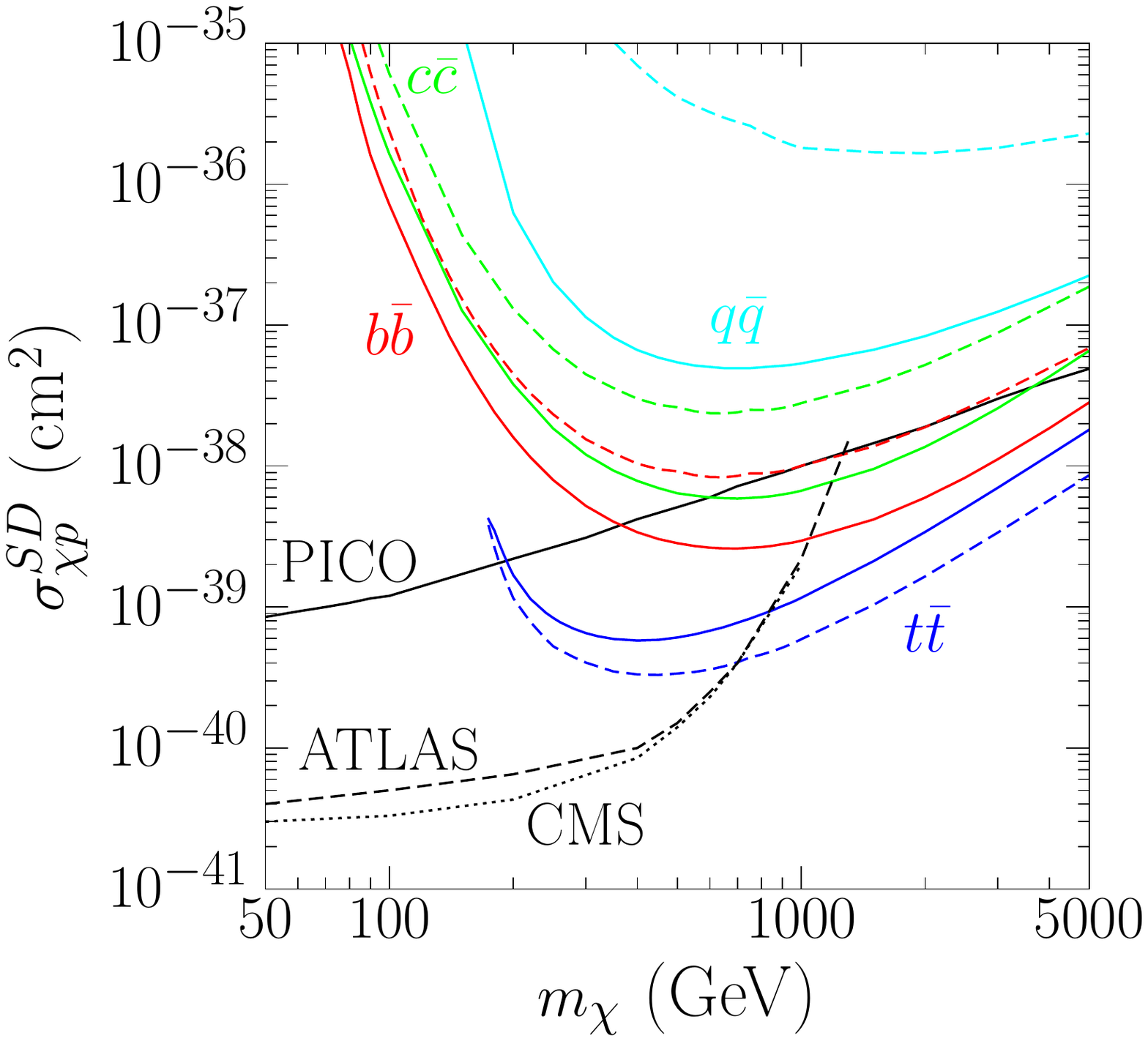}
\end{tabular}
\vspace{-2.8cm}
\caption[]{Upper limit on $\sigma^{SD}_{\chi p}$ for DM annihilation in $W^+W^-,ZZ,HH,\gamma\gamma$ (left panel) or  $b\bar{b},c\bar{c},t\bar{t},q\bar{q}$ (right panel) using the \pppc(full) or \wimpsim(dashed) spectra. The best limit from the DM direct detection  PICO~\cite{Amole:2015lsj} is also displayed as well as the limits from mono jet analysis in CMS~\cite{Khachatryan:2014rra} and ATLAS~\cite{Aad:2015zva}}
\label{fig:VV_quarks}
\end{center}
\end{figure}

Using the likelihood method described in section~\ref{sec:ns}, we can extract the limits on spin dependent interactions of DM on protons for each channel of DM annihilation into SM
particles for both the \pppc and \wimpsim spectra.
The results are displayed in Fig.~\ref{fig:leptons} and  Fig.~\ref{fig:VV_quarks}.

The strongest limits are obtained when DM annihilates into neutrinos and in particular into  $\nu_\mu$ or $\nu_{\tau}$. We find $\sigma^{SD}> 9.7\times 10^{-42}$~pb for  $m_\chi=250$~GeV using \wimpsim.
At masses above the TeV scale the $\nu_\tau$ channel becomes much more sensitive than the $\nu_\mu$ channel. This is due to the fact that $\nu_\tau$ can be regenerated, a charged current interaction of $\nu_\tau$ with nuclei generates a $\tau$ which in turns decay rapidly into $\nu_\tau$ with a high probability. On the other hand the $\mu$ generated from $\nu_\mu$ by charged current interactions is stopped in the Sun medium before emitting a neutrino while the electron generated from $\nu_e$ is stable.  Note that the different treatment of propagation in \wimpsim and \pppc leads to roughly a factor 2 in the 90\%CL limit on $\sigma^{SD}_{\chi p}$, with the \pppc leading to weaker limits because the neutrino spectrum is softer. For all masses above 60GeV the neutrino telescope limits are more stringent than the ones from direct detection. 

At high masses  the second most sensitive leptonic channel is  the $\tau$ channel, furthermore for all masses above 100 GeV sensitivity is better than direct detection. Note that DM annihilating into $\mu$ or electron pairs can also lead to a neutrino flux, even though muons are quickly absorbed in the Sun and electrons are stable. This is  due to electroweak bremstrahlung~\cite{Baratella:2013fya}. However the limits obtained are much weaker than for  the $\tau^+\tau^-$ channel. The effect of including lepton chirality is illustrated for the $\mu^+\mu^-$ channel and basically leads to an improvement on  the limit on SD cross section by a factor 2, this channel gives slightly more stringent limits than PICO in the range 600 -- 1000 GeV.

The results  for the $W$/$Z$ and Higgs channels are displayed in Fig.~\ref{fig:VV_quarks} (left panel). The most sensitive channels are respectively $ZZ$ and $W^+W^-$ for DM masses below the TeV scale, while above 2~TeV the channel $HH$ becomes slightly more sensitive. In fact at  masses of 5~TeV  the vector boson channels are within a factor 3 above the best leptonic channels.  Using  the \pppc tables  we have  extracted independently limits from the  $W_T$ and $W_L$ channels, the results are within 10\% of  those for the unpolarized case.  We have also computed the limits on $\sigma^{SD}$ assuming that DM annihilates solely into gluon or photon pairs, however they  are weak. We find  that for $m_\chi=800$~GeV,   $\sigma^{SD}> 6.6\times 10^{-38}$~pb in the gluon channel and $\sigma^{SD}> 1.9\times 10^{-38}$~pb in the photon channel.
When DM annihilates into quarks the best limits are obtained for heavy quarks, in particular from the $t\bar{t}$ channel which lies  about a factor of 3 above that of the $ZZ$ case.
In all those limits there is clearly a large uncertainty coming from the computation of the neutrino spectra as discussed above.  
Note that for light quarks it is the \pppc spectra that leads to the most stringent limits, as expected since in this case the neutrino
flux is larger, see Fig.~\ref{fig:comparison} for the $b\bar{b}$ channel.
 In fact for DM above 2 TeV, the $b\bar{b}$ channel has a comparable sensitivity (within a factor two) to the top or  boson channels. For both the third generation quarks the upper limit on $\sigma^{SD}_{\chi p}$ is better than the one obtained in direct detection for masses above a few hundred GeV's. 
For light DM masses, these limits are  however weaker than those obtained from the ATLAS and CMS collaborations in searches for DM in monojet events with 8 TeV data at the LHC~\cite{Aad:2015zva,Khachatryan:2014rra}. 
The collider limits presented in Fig.~\ref{fig:VV_quarks}  are based on  an effective field theory (EFT) approach and assume that DM couples to all quarks via the operator 
$1/M_*^2 \bar\chi\gamma^\mu\gamma^5\chi \bar{q}\gamma_\mu\gamma^5$. In this approach the production of DM at colliders can be directly related to the direct detection cross sections provided the scale of new physics inducing this operator is large enough for the EFT to be valid.  

\begin{figure}[hbt!]
\begin{center}
\hspace{-12mm}\includegraphics[scale=0.42]{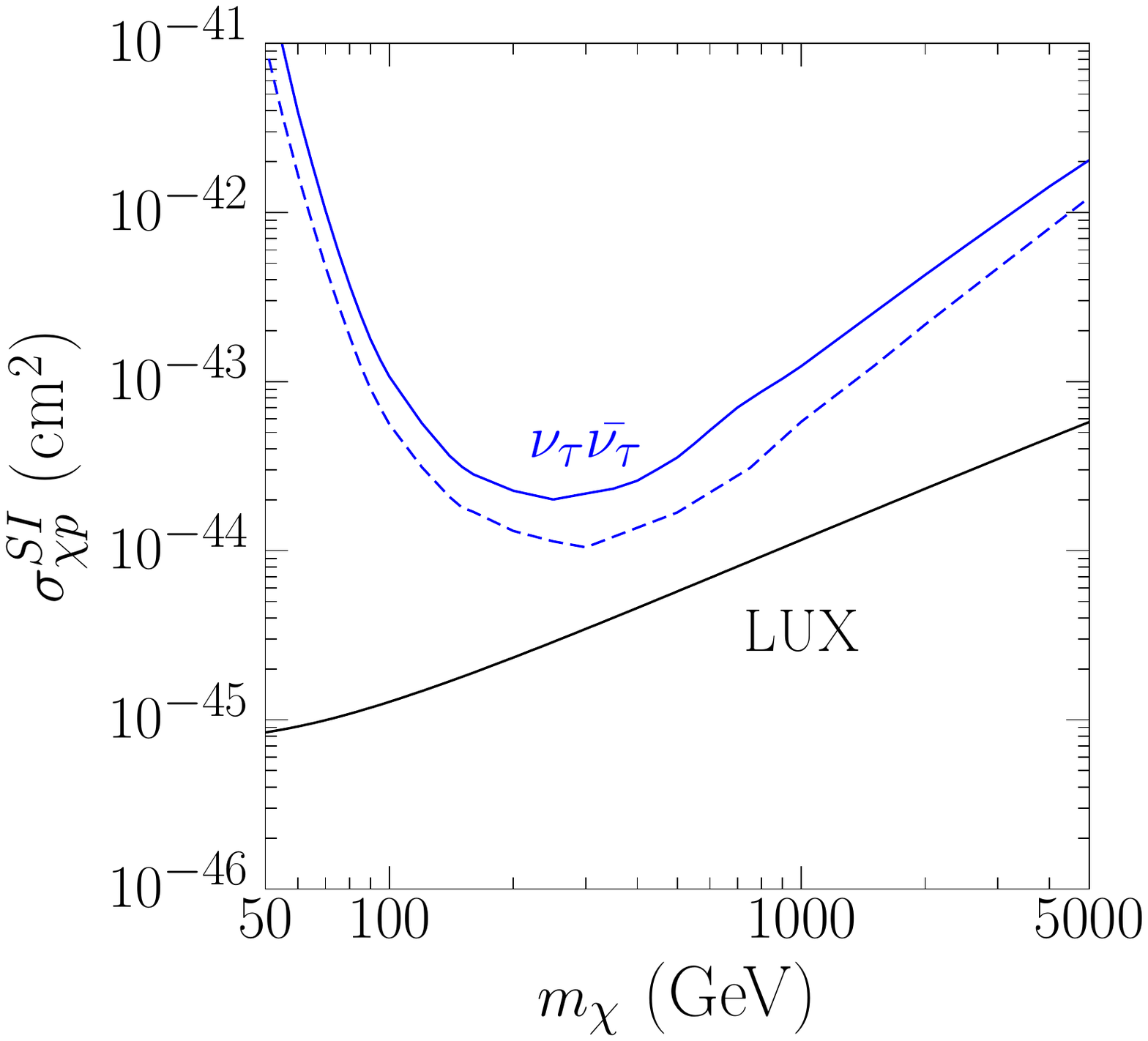}
\vspace{-2.2cm}
\caption[]{Upper limit on $\sigma^{SI}_{\chi p}$ for DM annihilation into $\nu_\tau\bar\nu_\tau$  using the  \pppc(full) or \wimpsim(dashed) spectra compared with 
 the best limit from DM direct detection in LUX~\cite{Akerib:2013tjd}.}
\label{fig:si_nu}
\end{center}
\end{figure}

We have also extracted the limits on $\sigma^{SI}$ assuming DM annihilates completely in $\nu_\tau$ pairs, here we assume equal cross section for scattering on protons or neutrons. We find that  even with this favorable channel, neutrino telescopes cannot compete with direct detection experiments, the best limit being a factor 2-10 above that of LUX~\cite{Akerib:2013tjd}, see Fig.~\ref{fig:si_nu}.

\section{Constraints on DM models}
\label{sec:DMmodels}

In a typical DM model, DM annihilation can proceed through different channels.
We investigate the constraints that can be derived from IceCube data within two sample DM models, a Z-portal model~\cite{Arcadi:2014lta}
and U(1) extensions of the MSSM with a neutralino DM~\cite{DaSilva:2013jga,Belanger:2015cra}.
Note that for this analysis, we include both SD and SI interactions and that we use the \wimpsim spectra.

\subsection{Z-portal dark matter}

This model is a simple extension of the SM with a fermion dark matter candidate which couples to the $Z$ with either vector ($V_\chi$) or axial-vector couplings ($A_\chi$),~\cite{Arcadi:2014lta}. 
Such interaction  can be obtained after symmetry breaking from the effective operator $ig/v^2 H^\dagger D_\mu H (\overline\chi\gamma^\mu (V_\chi+A_\chi\gamma_5)\chi)$ where $H$ is a scalar doublet.  The relevant DM interactions are described by the  Lagrangian 
\begin{equation}
{\cal L}=\frac{g}{4 \cos\theta_W} \overline\chi \gamma^\mu \left( V_\chi-A_\chi \gamma^5\right) \chi Z_\mu \left( 1+ \frac{g}{m_W} H  \right).
\label{eq:Zportal}
\end{equation}
In addition to the DM mass the model contains  only two free parameters, we choose $A_\chi$ and the ratio $\alpha=A_\chi/V_\chi$. 
 The  SD interaction which proceeds through $Z$ exchange depends only on the axial-vector coupling ($A_\chi$) while the SI interaction depends on the vector coupling ($V_\chi$). 
When kinematically accessible,  DM annihilation is mainly into $W$ pairs, which depends on $V_\chi$,  or into  final states involving $ZZ, ZH$ and heavy fermion pairs which mostly depend on $A_\chi$.
Annihilation into leptons and light quarks is generally suppressed. Note that terms involving the Higgs in Eq.~\eqref{eq:Zportal} are necessary to insure gauge invariance and have to be include in the relic density computation.
To assess the constraints originating from IceCube22, we have performed a random scan over the free parameters of the model  in the range
 
 \begin{equation}
 50 {\rm \, GeV} < m_\chi < 5000 {\rm \, GeV} \;\;\;\;   10^{-5}< |A_\chi|<1.0 \;\;\;\;   0.1 < |\alpha|< 10^5.
 \end{equation}

\begin{figure}[hbt!]
\begin{center}
\hspace{-12mm}\includegraphics[scale=0.42]{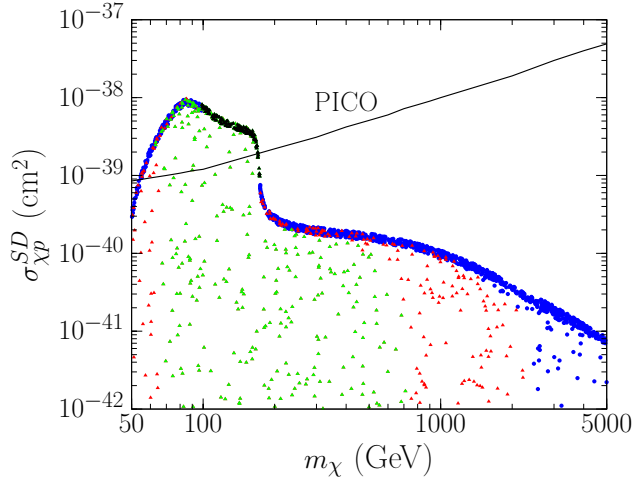}
\vspace{-2.2cm}
\caption[]{Constraints on the parameter space of the Z-portal model in the $\sigma^{SD}_{\chi p} - m_\chi$ plane. The color code indicates  points that  are constrained by IceCube22 and by LUX (green) by IceCube22 only (black) and by LUX only (red). Blue points satisfy all constraints and have a relic density in agreement with PLANCK. }
\label{fig:zdm}
\end{center}
\end{figure}
 
We have imposed the relic density constraint from PLANCK, allowing a 10\% uncertainty $0.9 < \Omega h^2/0.1199 < 1.1$~\cite{Planck:2015xua}. For the allowed points, we then determined those that are excluded by  LUX and/or  by IceCube, see Fig.~\ref{fig:zdm}. The former excludes points where $V_\chi$ is large while IceCube constrains points where $A_\chi$ is large. The DM mass is also relevant, 
for $m_{\rm DM}<m_W$ annihilation into light quarks dominates and this region is thus poorly constrained by IceCube.  In the region where  $m_W<m_{\rm DM}<m_{top}$ we  find the best IceCube exclusion since annihilation into  gauge bosons  leads to strong exclusion.  Heavier DM annihilates rather in $t\bar{t}$ or $ZH$, both have weaker  exclusion power than the $W^+W^-$ channel. For these masses it is generally LUX that excludes the points.
After applying the PLANCK constraints, for a given dark matter mass, most points in Fig.~\ref{fig:zdm} are concentrated in a narrow band for the SD cross section. This is somewhat an artifact of the scan, allowing very small values of $\alpha$ (or a dominant $V_\chi$ coupling) would populate the region below the band.  
However,  a large value for $V_\chi$  also leads to  a large spin independent cross section, thus all those points are constrained by the LUX result even if in some case they can be also constrained by IceCube. In summary after imposing the upper limit on SI interactions, IceCube can further constrain DM in the narrow mass range 80 -- 175~GeV.

\subsection{UMSSM}

U(1) extensions of the MSSM (UMSSM) are characterized by  additional vector and scalar superfields. The model therefore features a new gauge boson ($Z'$) and a singlet scalar, each with their supersymmetric partner, 
furthermore the model contains right-handed (RH) neutrinos. The lightest neutral supersymmetric particle can be the DM candidate, either a neutralino or the partner of one of the RH neutrino. Moreover new tree-level contributions to the Higgs mass make it possible to achieve 
a Higgs mass at 125 GeV without large corrections from the stop sector.  
We use  the results of a random scan performed in~\cite{Belanger:2015cra} where constraints from  flavor observables,  the Higgs mass and couplings, and  LHC searches for the $Z'$ and for supersymmetric particles were considered. 
Moreover an upper bound on the relic density of dark matter was imposed~\cite{Planck:2015xua}   as well as an upper bound from direct spin independent dark matter searches from LUX~\cite{Akerib:2013tjd}.
We then compute the CL for IceCube exclusion for each point of parameter space. In computing the capture rate we took into account the fact that the lightest neutralino $\tilde\chi^0_1$ could constitute only a fraction of the dark  matter using a rescaling factor $\xi$ defined by
\begin{eqnarray}
\xi = 
\begin{cases} 
\frac{\Omega_{\tilde\chi^0_1} h^2}{0.1168} \quad \, \textrm{for} \ \Omega_{\tilde\chi^0_1} h^2 <0.1168, \\   
1  \qquad \quad \textrm{for} \ \Omega_{\tilde\chi^0_1} h^2 \in [0.1168, 0.1208],
\end{cases}
\end{eqnarray}
where the $2\sigma$ deviation from the central value measured by PLANCK is $\Omega h^2 = 0.1168$.

\begin{figure}[hbt!]
\begin{center}
\hspace{-12mm}\includegraphics[scale=0.27]{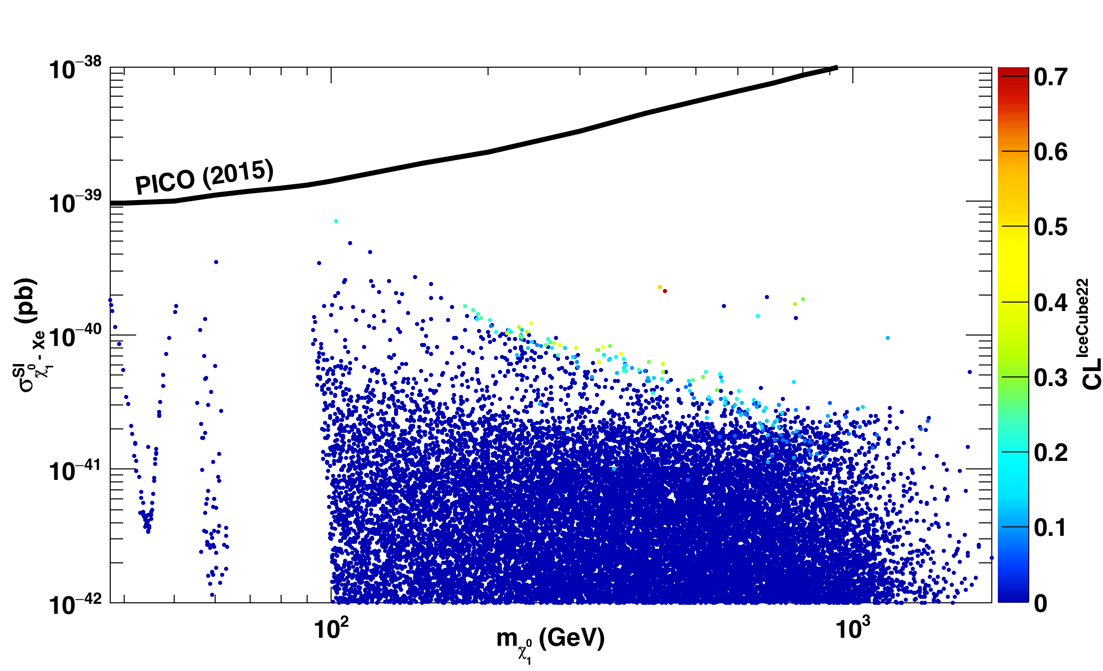}
\vspace{-0.1cm}
\caption[]{Allowed parameter space of the UMSSM model in the $\xi\sigma^{SD}_{\tilde\chi^0_1 - p} - m_{\tilde\chi^0_1}$ plane.}
\label{fig:umssm}
\end{center}
\end{figure}

Our results for all points satisfying the above mentioned constraints are shown in Fig.~\ref{fig:umssm} in  the spin dependent cross section vs neutralino mass plane. 
We did not find any point falling in the 90\%CL exclusion by IceCube22, although we found such points  when we ignored the LUX constraint.
Moreover when assuming some regeneration mechanism for DM, which implies ignoring the  rescaling of  the DM local density, a few points were excluded by IceCube. 
Since  our limits are too conservative at low masses we expect that IceCube79 data would exclude some of these points.
Indeed  a   prospective study  had shown that  in the MSSM several points were excludable by IceCube+DeepCore~\cite{Silverwood:2012tp}. Moreover after  applying the upper limit on the SI direct detection cross section such points are concentrated at masses below 250 GeV. 
Although the two studies are not directly comparable  since they are based on  different supersymmetric models and we  include most recent constraints form the LHC on the Higgs and supersymmetric sectors, it is encouraging that they qualitatively agree.

\section{Conclusion}
\label{sec:Conclusion}

We have shown that upper limits on spin dependent interactions of DM with protons obtained from IceCube22 data are stronger than those from direct detection in a variety of channels and for a large range of masses above 100 GeV.  The best limits are obtained for DM annihilation directly into neutrinos, although the $\tau^+\tau^-,W^+W^-,ZZ$ channels also display good sensitivity especially for masses of a few hundred GeVs . 
Uncertainties from the propagation and the oscillation of neutrinos in the Sun estimated by comparing the neutrino spectra computed by two different methods induce up to a factor two difference on the extracted limits on the DM proton spin dependent cross section. Other source of uncertainties affecting the limits derived from DM capture in the Sun arise from 
the DM local density, the DM velocity distribution,  the escape velocity in the Sun, and the presence of a dark disc. These  uncertainties were estimated to impact the capture rate by at most  50\%~\cite{Choi:2013eda}. 
Note however that direct detection limits are also sensitive to the uncertainties on the DM local density and velocity distribution.

A direct comparison of the  LHC  and neutrino telescope limits on the spin dependent cross section is straightforward in an EFT approach when we assume that DM couples to quarks only.    In this case we found  that the collider limits were better for light quarks and for light DM. 

We have also shown that the IceCube22 results constrain some area of the parameter space of typical dark matter models, although the potential is much more limited when the current constraints on the DM spin independent scattering cross section on nucleons are taken into account first.

The results presented here and implemented in {\tt micrOMEGAs\_4.2} are based on the publicly available data from IceCube22, the \micro facilities to compute limits on DM proton cross section from neutrino telescopes will be extended when more recent IceCube data becomes publicly available.

\section{Acknowledgements}
We thank Marco Cirelli, Joakim Edsj\"{o} and Antje Putze for useful communication.
 This work was  supported in part by the LIA-TCAP of CNRS,  by the French ANR, Project DMAstro-LHC, ANR-12-BS05-0006, by the {\it Investissements d'avenir}, Labex ENIGMASS and by the European Union as part of the FP7 Marie Curie Initial Training Network MCnetITN (PITN-GA-2012-315877).
 The work of AP  was  also supported by the Russian foundation for Basic Research, grant RFBR-15-52-16021-CNRS-a.

\begin{appendices}

\section{micrOMEGAs functions for neutrino telescopes}

Details on all \micro functions can be found in the manual contained in the \verb|man| directory of each version. 
Here we present the functions to compute the (anti-)neutrino fluxes as well as the likelihood and the CL corresponding to the data of Icecube22  for DM nucleon cross sections
defined  in a model-independent way or within the context of a specific model. 

We have introduced a new global parameter {\tt WIMPSIM} which allows to
choose  the   neutrino spectra. The default value  {\tt WIMPSIM=0} 
corresponds to the \pppc\cite{Baratella:2013fya} spectra while  {\tt WIMPSIM=1} corresponds to the
\wimpsim \cite{Blennow:2007tw} spectra and  {\tt WIMPSIM=-1}  to the 
 \pppcold\cite{Cirelli:2005gh} spectra implemented in previous~\micro versions. 
The parameter {\tt forSun=1} is used to compute the signature of DM annihilation in the Sun and {\tt Mdm} designates the  mass of the DM particle. \\
\noindent
$\bullet$ \verb|basicNuSpectra(forSun,Mdm,pdg, pol, nu_spect, nuB_spect) |\\
calculates  the $\nu_{\mu}$ and  $\bar{\nu}_{\mu}$ spectra corresponding to one DM annihilation 
into a particle-antiparticle pair with PDG code {\tt pdg}. \verb|Mdm| is the DM mass.
Note that this
routine depends implicitly on the global parameter {\tt WIMPSIM}.
The parameter {\tt pol} selects the spectra for  polarized particles available  in
\pppc. {\tt pol=-1(1)} corresponds to longitudinal (transverse) polarisation of vector bosons or to left-handed (right-handed) fermions, {\tt pol=0}
is used for unpolarized spectra. When polarized spectra are not available, the unpolarized ones are generated irrespective of the value of \verb|pol|. 
The resulting spectrum is stored in arrays {\tt nu\_spect} and {\tt nuB\_spect}.

\noindent $\bullet$ \verb|captureAux(fv,forSun,Mdm,csIp,csIn,csDp,csDn)|\\
calculates the number of DM particles captured per second assuming the cross sections
for  spin independent and spin dependent 
interactions with protons and neutrons   {\tt csIp, csIn, csDp, csDn} are
given as input parameters (in {\tt [pb]}). 
A negative value for one of the cross sections  is interpreted as a destructive
interference between the  proton and neutron amplitudes. The first argument is the DM velocity distribution, 
usually we substitute {\verb|fv = Maxwell|}. The result  depends implicitly on the global parameters {\tt rhoDM}  representing the local DM density,  by default {\tt rhoDM} $=0.3\,\rm{GeV}/\rm{cm}^3$.

\noindent
$\bullet$ \verb|IC22nuAr(E)|\\
effective area in [$\rm{km}^2$] as a function of the neutrino energy, $A_{\nu_{\mu}}(E)$.

\noindent$\bullet$ \verb|IC22nuBarAr(E)|\\
effective area in [$\rm{km}^2$] as a function of the anti-neutrino energy, $A_{\bar{\nu}_{\mu}}(E)$.

\noindent
$\bullet$ \verb|spectrMult(Spec, func)|\\
allows to multiply the spectrum \verb|Spec| by any energy dependent function \verb|func|.

\noindent
$\bullet$ \verb|spectrInt(E1,E2,Spec) | \\
integrates a spectrum/Flux, \verb|Spec| from $E_1$ to  $E_2$.\\

\noindent
$\bullet$ \verb|IC22BGdCos(cs)| \\
angular distribution of the number of background events as a function of $\cos\varphi$, $\frac{dN_{bg}}{d\cos{\varphi}}$.
  
\noindent 
$\bullet$ \verb|IC22sigma(E)|\\
neutrino angular resolution in radians  as a function of energy,
Eq.~\eqref{eq:N_cosphi}.

\noindent
$\bullet$ \verb|exLevIC22( nu_flux, nuB_flux,&B)|\\
 calculates the credibility level for number of signal events  generated by   given $\nu_{\mu}$ and  $\bar{\nu}_{\mu}$  fluxes, Eq.~\eqref{eq:pvalue}. The fluxes are assumed to be in
$[\rm{GeV\;km}^2\;\rm{Year}]^{-1}$.  This function uses
the \verb|IC22BGdCos(cs)| and \verb|IC22sigma(E)| angular distribution for background and signal as well as the event files distributed  by IceCube22 
with  $\varphi<\varphi_{cut}=8^{\circ} $.  The returned parameter \verb|B|  corresponds to  the Bayes factor, ${\cal L}(1,\varphi_{cut})/{\cal L}(0,\varphi_{cut})$, 
with $\cal L$ defined in Eq.~\eqref{eq:likelihood}.\\

\noindent  
$\bullet$ \verb|fluxFactorIC22(exLev, nu,nuBar)|\\
For given neutrino, \verb|nu|, and anti-neutrino, \verb|nuBar|,  fluxes, 
this function returns the factor that should be applied to the 
fluxes  to obtain a given exclusion
level {\tt exLev}  in exLevIC22. This is used to obtain  limits on the SD cross
section for given annihilation channel.

\noindent
$\bullet$ \verb|IC22events(nu,nuB, phi, &Nsig,&Nbg,&Nobs)|\\
For given neutrino and anti-neutrino fluxes  this routine calculates the expected number of signal
{\tt Nsig}, background {\tt Nbg} and observed  {\tt Nobs} events. 
\verb|phi| is the angular cut defined in degree units. The last
parameter has to be an {\tt int} type variable. Note that in 
calculating exclusion limits on cross sections,  a factor 1.2 to take into account the uncertainty on the effective area of the
detector is included.

To calculate the neutrino fluxes within some theoretical DM model one can use instead of \verb|basicNuSpectra| and \verb|captureAux|, the function\\ 
$\bullet$ \verb| neutrinoFlux(fv,forSun,nu, nu_bar)|\\
which calculates  the muon neutrino/anti-neutrino  fluxes  near the surface of the Earth. 
This function  a)  calculates the capture,   annihilation, and  evaporation rates
in Eq.~\eqref{eq:ndot} and b) solves it  numerically;
c) calculates all branchings of DM annihilation and substitute them in Eq.~\eqref{eq:nu_spectra} to get the fluxes.
Here  \verb|fv|  is  the DM velocity distribution normalized such that $\int_o^\infty v f(v) dv=1$.
The units  are $km/s$ for v and $s^2/km^2$ for  
f(v).  For example one can use  the  \verb|Maxwell| function.   The calculated fluxes are stored in the arrays {\tt nu} and {\tt nu\_bar}, the units used are $\rm{[GeV\;km}^2\;\rm{Year}]^{-1}$. \\

In~\micro the neutrino spectra and fluxes are tabulated in arrays of 
{\it double} numerical type with  $NZ=250$ elements. The first (zeroth in C)
element of this array contains the maximal  energy of the spectrum.  It usually corresponds to the
DM mass, see   \cite{Belanger:2013oya}.  The functions to interpolate and display spectra include

\noindent
$\bullet$ \verb|SpectdNdE(E,Flux)|\\ interpolates the tabulated spectra/flux and returns energy distribution.

\noindent
$\bullet$ \verb|displaySpectra(title, Emin, Emax, N, nu1,lab1,...)|\\
displays several spectra. Here {\tt title} contains some text, {\tt Emin,Emax} are the lower and upper limits, and {\tt N} is the number of spectra to display.
Each spectrum is defined with  two arguments, \verb|nu1| designates the spectrum array and \verb|lab1| contains  some text to label the spectrum.

\section{Examples}

An example of the code that computes  the 90\% exclusion  limits for
DM-proton spin dependent cross sections  for  $W$  annihilation channels
and reproduces data of Tables~\ref{tab:signalToCS}, \ref{tab:icecube_new_W} and \ref{tab:icecube_new_B} reads :

\begin{footnotesize}
\begin{verbatim}
 double  nu[NZ],nuB[NZ]; // arrays for neutrino and antineutrino spectra 
 double  csSDp[6]={1.E-40,9.51E-41,1.04E-40, 2.67E-40,2.06e-39,6.0e-39};
                         // cross section TAB_1,4-th column 
 double  dSun=150E6;     // distance to Sun [km] 
 double  yrs=31556925.2; // year [sec] 
 double  Mdm[6]={100,250,500,1000,3000,5000}; // DM masses 
 int forSun=1;
 double exLev=0.9; // 90% exclusion confidence level 
 double Nsig;      // for number of signal events 
             
 WIMPSIM=1;
 printf("W channel:\n");

 for(int k=0;k<6;k++)   
 {  
    double Crate=captureAux(Maxwell,forSun,Mdm[k],0,0,csSDp[k]*1E36,0); //DM capture rate[s]
    double gamma=Crate*yrs/(4*M_PI*dSun*dSun)/2;
    double f_90; // improving factor for 90% confidence level exclusion 
    printf("Dark matter mass is %.2E\n",Mdm[k]);
    basicNuSpectra(forSun,Mdm[k],24/*W*/, 0, nu, nuB);              // neutrino spectra 
    for(int i=1;i<NZ;i++) { nu[i]*=gamma; nuB[i]*=gamma;}           // neutrino fluxes 
    if(k)
    {  IC22events(nu,nuB, 10, &Nsig,NULL, NULL);
       printf("     csSDp=%.2E[cm^2] ==>  Nsignal=%.2E (TAB_1 column 1)\n",csSDp[k],Nsig);
    } 
    f_90=fluxFactorIC22( exLev,nu,nuB);  
    printf("     90%%  exclusion limit for csSDp is %.2E[cm^2]\n",csSDp[k]*f_90);
    for(int i=1;i<NZ;i++) {nu[i]*=f_90; nuB[i]*=f_90;}
    IC22events(nu,nuB, 10,&Nsig,NULL,NULL);
    printf("     90%% exclusion limit for number of signal events: %.2E \n",Nsig/1.2);
 } 
\end{verbatim}
\end{footnotesize}

The code for the calculation of an exclusion level in the framework of a given model
reads

\begin{footnotesize}
\begin{verbatim}
  double nu[NZ],nuB[NZ]; 
  int forSun=1;
  WIMPSIM=1;
  neutrinoFlux(Maxwell,forSun,nu,nuB);
  printf("IceCube22 exclusion confidence level = %.2E%%\n",100*exLevIC22(nu,nu_bar,NULL));
\end{verbatim}
\end{footnotesize}
For the default data file  {\tt mssmh.par} provided in the  MSSM directory of micrOMEGAs,  we get a  confidence level 
of 18\%.

\section{\darksusy} 
\label{usageDS}
\darksusy is generally used in the framework of some SUSY model.  In order to compare the results obtained with \micro in a model independent approach
we use the following trick  to  calculate the number of signal events  for a  specific channel and a 
DM -- nucleon cross section.
The \darksusy subroutine {\tt dsntICbounds} provides the number of signal events expected in IceCube22 for given annihilation channels specified by the array {\it wabr}, for a DM mass {\it wamwimp}, and for  DM – nucleon cross sections with the parameters {\it wasigsdp} for spin dependent and {\it wasigsip} for spin independent. We therefore redefine these parameters stored in {\tt  common /wabranch/} before the call to the subroutine {\tt dsntICbounds}.
 All  cross sections should be specified in  [$\rm{cm}^2$].
 The content of the  array {\it  wabr} is given   in the file 
{\tt DarkSUSY/src/wa/dswayieldone.f}. In particular {\it wabr(13)} corresponds to the  $W^+W^-$ annihilation channel and 
 {\it wabr(25)} to the $ b\bar{b}$ channel.
 
\end{appendices}

\bibliography{ice22_micro}
\end{document}